\def\lb{\hfil\break}
\def\pmb#1{\setbox0=\hbox{#1}%
   \kern-.025em\copy0\kern-\wd0
   \kern.05em\copy0\kern-\wd0
   \kern-0.025em\raise.0433em\box0}
\def\gta{\mathrel{{\lower 3pt\hbox{$\mathchar"218$}}\hskip-8pt
   \raise 2pt\hbox{$\mathchar"13E$}}}
\def\lta{\mathrel{{\lower 3pt\hbox{$\mathchar"218$}}\hskip-8pt
   \raise 2pt\hbox{$\mathchar"13C$}}}
\def\half{{\scriptstyle{1\over2}}}
\def\dagg{\phantom{\dagger}}            
\def\subboldc{\pmb{$\scriptstyle c$}}   
\def\boldd{\pmb{d}}                     
\def\boldell{\pmb{$\ell$}}              
\def\subboldell{\pmb{$\scriptstyle\ell$}}
\def\boldT{\pmb{$T$}}                   
\def\bolddelta{\pmb{$\delta$}}
\def\subbolddelta{\pmb{$\scriptstyle\delta$}}
\def\boldeta{\pmb{$\eta$}}
\def\subboldeta{\pmb{$\scriptstyle\eta$}}
\def\boldxi{\pmb{$\xi$}}
\def\vacuum{|\pmb{\it O}\thinspace\rangle}
\def\bfnabla{\bf \pmb{$\nabla$}\/}
\def\bfcalE{\bf \pmb{${\cal E}$}\/}
\def\up{\uparrow}
\def\dn{\downarrow}
\def\ud{\uparrow\downarrow}
\def\today{\number\day\space\ifcase\month\or
  January\or February\or March\or April\or May\or June\or
  July\or August\or September\or October\or November\or December\fi
 \space\number\year}
\documentstyle[twocolumn,aps]{revtex}
\font\tenrm=cmr8
\begin{document}
\newcommand{\Vv }{{\raisebox{-1.2pt}{\makebox(0,0){$o$}}}}
\newcommand{\Zz }{{\raisebox{-1.2pt}{\makebox(0,0){$\mbox{\tiny o}$}}}}
\newcommand{\Xx }{{\special{em:moveto}}}
\newcommand{\Yy }{{\special{em:lineto}}}
\newcommand{\Ww }{{\usebox{\plotpoint}}}
\title{\begin{minipage}{6.5in}
 {\large \ \\ \ \\ \ \\  \ \\
\centerline{STRIPE FLUCTUATIONS, CARRIERS, SPECTROSCOPIES, TRANSPORT,}
\centerline{AND BCS--BEC CROSSOVER IN THE HIGH-$T_c$ CUPRATES} }
{\normalsize  \ \\ 
\centerline{J. ASHKENAZI}
\centerline{Physics Department, University of Miami, P.O. Box 248046,
Coral Gables, FL 33124, U.S.A.} \\ \ \\ \ } 
\end{minipage}
}
\author{\ 
\begin{minipage}{5.125in}
\marginparwidth 0.625in 
\small
\baselineskip 9pt
{\bf Abstract}---The quasiparticles of the high-$T_c$ cuprates are found
to consist of: polaron-like ``stripons'' carrying charge, and associated
primarily with large-$U$ orbitals in stripe-like inhomogeneities;
``quasi-electrons'' carrying charge and spin, and associated with
hybridized small-$U$ and large-$U$ orbitals; and ``svivons'' carrying
spin and lattice distortion. It is shown that this electronic structure
leads to the systematic behavior of spectroscopic and transport
properties of the cuprates. High-$T_c$ pairing results from transitions
between pair states of stripons and quasi-electrons through the exchange
of svivons. The cuprates fall in the regime of crossover between BCS and
preformed-pairs Bose-Einstein condensation behaviors. 
\end{minipage}
}
\maketitle
\setlength{\unitlength}{1in}
\makeatletter
\global\@specialpagefalse
\def\@oddhead{\footnotesize \it \hfill \ Journal of Physics and 
Chemistry of Solids}
\makeatother
\baselineskip 12pt
\normalsize \rm
%

First-principles calculations in the cuprates support an approach based
on the existence of both ``large-$U$'' and ``small-$U$'' orbitals in the
vicinity of the Fermi level ($E_{_{\rm F}}$). Let us denote the fermion
creation operator of a small-$U$ electron in band $\nu$, spin $\sigma$
(which can be assigned a number $\pm 1$), and wave vector ${\bf k}$ by
$c_{\nu\sigma}^{\dagger}({\bf k})$. The large-$U$ orbitals in the
CuO$_2$ planes are approached through the ``slave-fermion'' method [1].
A large-$U$ electron in site $i$ and spin $\sigma$ is then created by
$d_{i\sigma}^{\dagger} = e_i^{\dagger} s_{i,-\sigma}^{\dagg}$, if it is
in the ``upper-Hubbard-band'', and by $d_{i\sigma}^{\prime\dagger} =
\sigma s_{i\sigma}^{\dagger} h_i^{\dagg}$, if it is in a Zhang-Rice-type
``lower-Hubbard-band''. Here $e_i^{\dagg}$ and $h_i^{\dagg}$ are
(``excession'' and ``holon'') fermion operators, and
$s_{i\sigma}^{\dagg}$ are (``spinon'') boson operators. These auxiliary
particles have to satisfy the constraint: $e_i^{\dagger} e_i^{\dagg} +
h_i^{\dagger} h_i^{\dagg} + \sum_{\sigma} s_{i\sigma}^{\dagger}
s_{i\sigma}^{\dagg} = 1$. 

Within the ``auxiliary space'' a chemical-potential-like Lagrange
multiplier is introduced to impose the constraint on the average.
Physical observables are calculated as combinations of Green's functions
of the auxiliary space. Since the time evolution of Green's functions is
determined by the Hamiltonian which obeys the constraint rigorously, it
is not expected to be violated as long as the approximations used for
these Green's functions are justifiable. 


The spinon states are diagonalized by applying the Bogoliubov
transformation [2]: 
\begin{equation}
s_{\sigma}^{\dagg}({\bf k}) = \cosh{(\xi_{{\bf k}})}
\zeta_{\sigma}^{\dagg}({\bf k}) + \sinh{(\xi_{{\bf k}})}
\zeta_{-\sigma}^{\dagger}(-{\bf k}). 
\end{equation} 
The Bose operators $\zeta_{\sigma}^{\dagger}({\bf k})$ create spinon
states of ``bare'' energies $\epsilon^{\zeta} ({\bf k})$ which have a
V-shape zero minimum at ${\bf k}={\bf k}_0$. Bose condensation results
in antiferromagnetic (AF) order at wave vector ${\bf Q}=2{\bf k}_0$. The
values of ${\bf k}$ are within a Brillouin zone (BZ) determined by
adding ${\bf Q}$ to the basic reciprocal lattice. For ${\bf k} \to {\bf
k}_0$ one has [2] $\cosh{(\xi_{{\bf k}})} \cong -\sinh{(\xi_{{\bf k}})}
\gg 1$. 

Theoretical calculations [3] predict that a lightly doped AF plane tends
to form a frustrated striped structure where narrow charged stripes form
antiphase domain walls separating wider AF stripes. Various experiments
[4] (including EXAFS, magnetic neutron scattering, channeling, inelastic
neutron scattering, NQR, NMR, ARPES, PDF, heat transport, phonon
screening, {\it etc}) support the idea that the high-$T_c$ cuprates are
characterized by such a fluctuating striped structure. Within the one
dimensional charged stripes the spin-charge separation approximation is
expected to be valid. Under this approximation two-particle spinon-holon
(spinon-excession) Green's functions are decoupled into
single-auxiliary-particle Green's functions, and auxiliary particles can
be interpreted as physical quasiparticles. 

Holons (excessions) within the charged stripes are referred to as
``stripons''. Their creation operators (creating charge $-{\rm e}$) are
denoted by $p^{\dagger}_{\mu}({\bf k})$, and bare energies by
$\epsilon^p_{\mu}({\bf k})$. Since these dynamical stripes are highly
inhomogeneous and segmented, we find it more appropriate to assume a
starting point of localized, rather than itinerant, stripon states. The
${\bf k}$ quantum number here presents a ${\bf k}$-symmetrized
combination of degenerate localized states to be treated in a
perturbation expansion. These values of ${\bf k}$ are within a BZ based
on periodic supercells which are large enough to approximately contain
(each) the entire spectrum $\epsilon^p_{\mu}$ of bare stripon energies.
These supercells introduce an approximate long-range order in spite of
the local inhomogeneity introduced by the dynamical striped structure. 

Holons (excessions) away from the charged stripes are not decoupled from
the spinons. One can construct within the auxiliary space approximate
fermion creation operators of coupled holon-spinon (excession-spinon)
basis states as follows: 
\begin{eqnarray}
f_{\lambda\sigma}^{\dagger}({\bf k}^{\prime}, {\bf k}) &=&
{e_{\lambda}^{\dagger}({\bf k}^{\prime})
s_{\lambda,-\sigma}^{\dagg}({\bf k}^{\prime} - {\bf k}) \over
\sqrt{n^e_{\lambda}({\bf k}^{\prime}) + n^s_{\lambda,-\sigma}({\bf
k}^{\prime} - {\bf k})}}, \\ 
g_{\lambda\sigma}^{\dagger}({\bf k}^{\prime}, {\bf k}) &=& {\sigma
h_{\lambda}^{\dagg}({\bf k}^{\prime}) s_{\lambda\sigma}^{\dagger}({\bf
k} - {\bf k}^{\prime}) \over \sqrt{n^h_{\lambda}({\bf k}^{\prime}) +
n^s_{\lambda\sigma}({\bf k} - {\bf k}^{\prime})}}, 
\end{eqnarray} 
where these values of ${\bf k}$ are within the BZ of the basic lattice.
An index $\lambda$ has been introduced to account for the effect of more
than one CuO$_2$ layer within the unit cell, and: $n^e_{\lambda}({\bf
k}) \equiv \langle e_{\lambda}^{\dagger}({\bf k})
e_{\lambda}^{\dagg}({\bf k}) \rangle$, $n^h_{\lambda}({\bf k}) \equiv
\langle h_{\lambda}^{\dagger}({\bf k}) h_{\lambda}^{\dagg}({\bf k})
\rangle$, $n^s_{\lambda\sigma}({\bf k}) \equiv \langle
s_{\lambda\sigma}^{\dagger}({\bf k}) s_{\lambda\sigma}^{\dagg}({\bf k})
\rangle$. 

The mean-field eigenstates of the Hamiltonian, obtained within the
auxiliary space as combinations of the above basis states (more
rigorously, these states have to be orthogonalized to the stripon
states, and depleted), and the small-$U$ states [created by
$c_{\nu\sigma}^{\dagger}({\bf k})$] are referred to as
``quasi-electron'' (QE) states. Their creation operators are denoted by
$q_{\iota\sigma}^{\dagger}({\bf k})$, and bare energies by
$\epsilon^q_{\iota} ({\bf k})$. These energies form quasi-continuous
ranges of bands within the BZ, and part of them cross $E_{_{\rm F}}$.
The QE states close to $E_{_{\rm F}}$ introduce fluctuations to the AF
stripes, but do not destroy the AF correlations. 


The QE, stripon and spinon fields are strongly coupled to each other due
to hopping and hybridization terms of the Hamiltonian. This coupling can
be expressed through an effective Hamiltonian term whose parameters can
be in principle derived self-consistently from the original Hamiltonian.
Discussing, {\it e.g.}, the case of p-type cuprates, the coupling
Hamiltonian has the form: 
\begin{eqnarray}
{\cal H}^{\prime} &=& {1 \over \sqrt{N}} \sum_{\iota\lambda\mu\sigma}
\sum_{{\bf k}, {\bf k}^{\prime}} \big\{\sigma
\epsilon^{qp}_{\iota\lambda\mu}({\bf k}, {\bf k}^{\prime})
q_{\iota\sigma}^{\dagger}({\bf k}) p_{\mu}^{\dagg}({\bf k}^{\prime})
\nonumber \\ &\ &\times [\cosh{(\xi_{\lambda,{\bf k} - {\bf
k}^{\prime}})} \zeta_{\lambda\sigma}^{\dagg}({\bf k} - {\bf k}^{\prime})
\nonumber \\ &\ &+ \sinh{(\xi_{\lambda,{\bf k} - {\bf k}^{\prime}})}
\zeta_{\lambda,-\sigma}^{\dagger}({\bf k}^{\prime} - {\bf k})] + h.c.
\big\}, 
\end{eqnarray} 
where the ${\bf k}$ values correspond to the supercell stripons BZ,
within which the other fields have been embedded, multiplying the number
of their bands, and redefining their band indices appropriately. Thus
${\cal H}^{\prime}$ introduces a vertex between the QE, stripon and
spinon propagators [5].

As was observed [6] a localized stripon modifies the lattice in its
vicinity. Thus, any physical process induced by ${\cal H}^{\prime}$,
where a stripon is transformed into a QE, or vice versa, and a spinon is
emitted and/or absorbed, necessarily involves also the emission and/or
absorption of phonons. This can be formulated by multiplying a spinon
propagator, linked to the ${\cal H}^{\prime}$ vertex with a power series
of phonon propagators [5]. We refer to such a phonon-``dressed" spinon
as a ``svivon"; it carries spin and lattice distortion. The ${\cal
H}^{\prime}$ vertex is now interpreted as coupling between a QE,
stripon, and svivon propagators. 

The electrons spectral function at momentum ${\bf p}$ and energy
$\omega$, $A_e({\bf p}, \omega) \equiv \Im {\cal G}_e({\bf p},
\omega-i0^+) / \pi$ (where ${\cal G}_e$ is the electrons Green's
function) is expressed in terms of auxiliary space spectral functions
$A^q_{\iota}({\bf k}, \omega)$, $A^p_{\mu}({\bf k}, \omega)$, and
$A^{\zeta}_{\lambda}({\bf k}, \omega)$ of the QE's, stripons, and
svivons, respectively. It will be shown below that the resulting stripon
bandwidth is much smaller than the QE and svivon bandwidths. Thus, the
same phase-space argument as in the Migdal theorem can be applied to
conclude that vertex corrections to the ${\cal H}^{\prime}$ vertex are
negligible. Consequently, a second-order perturbation expansion in
${\cal H}^{\prime}$ is applicable calculating self-energy correction
(see the diagrams in Ref. [5]). The following expressions are obtained
for the QE, stripon, and svivon scattering rates $\Gamma^q({\bf k},
\omega)$, $\Gamma^p({\bf k}, \omega)$, and $\Gamma^{\zeta}({\bf k},
\omega)$ [$\Gamma({\bf k}, \omega) \equiv 2\Im \Sigma({\bf k},
\omega-i0^+)$]: 
\begin{eqnarray}
\Gamma&^q_{\iota\iota^{\prime}}&({\bf k}, \omega) \cong {2\pi \over N}
\sum_{\lambda\mu{\bf k}^{\prime}} \int d\omega^{\prime}
\epsilon^{qp}_{\iota\lambda\mu}({\bf k}^{\prime}, {\bf
k})\epsilon^{qp}_{\iota^{\prime}\lambda\mu}({\bf k}^{\prime}, {\bf k})^*
\nonumber \\ &\times& A^p_{\mu}({\bf k}^{\prime}, \omega^{\prime})
[-\cosh{^2(\xi_{\lambda,{\bf k} - {\bf k}^{\prime}})}
A^{\zeta}_{\lambda}({\bf k} - {\bf k}^{\prime}, \omega -
\omega^{\prime}) \nonumber \\ &+& \sinh{^2(\xi_{\lambda,{\bf k} - {\bf
k}^{\prime}})} A^{\zeta}_{\lambda}({\bf k} - {\bf k}^{\prime},
\omega^{\prime} - \omega)] \nonumber \\ &\times&
[f_{_T}(\omega^{\prime}) + b_{_T}(\omega^{\prime} - \omega)], \\ 
\Gamma&^p_{\mu\mu^{\prime}}&({\bf k}, \omega) \cong {2\pi \over N}
\sum_{\iota{\bf k}^{\prime}\sigma} \int d\omega^{\prime}
\epsilon^{qp}_{\iota\lambda\mu}({\bf k}^{\prime}, {\bf k})^*
\epsilon^{qp}_{\iota\lambda\mu^{\prime}}({\bf k}^{\prime}, {\bf k})
\nonumber \\ &\times& A^q_{\iota}({\bf k}^{\prime}, \omega^{\prime})
[\cosh{^2(\xi_{\lambda,{\bf k}^{\prime} - {\bf k}})}
A^{\zeta}_{\lambda}({\bf k}^{\prime} - {\bf k}, \omega^{\prime} -
\omega) \nonumber \\ &-& \sinh{^2(\xi_{\lambda,{\bf k}^{\prime} - {\bf
k}})} A^{\zeta}_{\lambda}({\bf k}^{\prime} - {\bf k}, \omega -
\omega^{\prime})] \nonumber \\ &\times& [f_{_T}(\omega^{\prime}) +
b_{_T}(\omega^{\prime} - \omega)], \\ 
\Gamma&^{\zeta}_{\lambda\lambda^{\prime}}&({\bf k}, \omega) \cong {2\pi
\over N} \sum_{\iota{\bf k}^{\prime}\mu} \int d\omega^{\prime}
\epsilon^{qp}_{\iota\lambda\mu}({\bf k}^{\prime}, {\bf k}^{\prime} -
{\bf k})^* \nonumber \\ &\times&
\epsilon^{qp}_{\iota\lambda^{\prime}\mu}({\bf k}^{\prime}, {\bf
k}^{\prime} - {\bf k}) [\cosh{(\xi_{\lambda{\bf k}})}
\cosh{(\xi_{\lambda^{\prime}{\bf k}})} A^q_{\iota}({\bf k}^{\prime},
\omega^{\prime}) \nonumber \\ &\times& A^p_{\mu}({\bf k}^{\prime} - {\bf
k}, \omega^{\prime} - \omega) + \sinh{(\xi_{\lambda{\bf k}})}
\sinh{(\xi_{\lambda^{\prime}{\bf k}})} \nonumber \\ &\times&
A^q_{\iota}({\bf k}^{\prime}, -\omega^{\prime}) A^p_{\mu}({\bf
k}^{\prime} - {\bf k}, \omega - \omega^{\prime})] \nonumber \\ &\times&
[f_{_T}(\omega^{\prime} - \omega)- f_{_T}(\omega^{\prime})], 
\end{eqnarray} 
where $f_{_T}(\omega)$ and $b_{_T}(\omega)$ are the Fermi and Bose
distribution functions at temperature $T$. 

The real parts of the self energies and the spectral functions are
obtained through the relations: 
\begin{eqnarray}
\Re\Sigma({\bf k}, \omega) &=& \wp\int {d\omega^{\prime} \Gamma({\bf k},
\omega^{\prime}) \over 2\pi (\omega - \omega^{\prime})}, \\ 
A({\bf k}, \omega) &=& {\Gamma({\bf k}, \omega)/2\pi \over [\omega -
\epsilon({\bf k}) - \Re\Sigma({\bf k}, \omega)]^2 + [\Gamma({\bf k},
\omega)/2]^2}. 
\end{eqnarray} 
Matrix diagonalization is necessary to determine the poles of the 
Green's functions, however only the diagonal terms are needed to
determine the spectral functions. 

For sufficiently doped cuprates a self-consistent solution is obtained
with three energy ranges. Expressions are derived below for the
intermediary energy range. The high energy range, above few tenths of an
eV (determined by the exchange energies of the large-$U$ spins) is
treated by introducing cut-off integration limits at $\pm\omega_c$ to
the Eq.~(8) integrals, resulting in spurious logarithmic divergences at
$\pm\omega_c$. The low energy range, below $\sim 0.02\;$eV, introduces a
``zero-energy'' non-analytic behavior to these expressions (analyticity
is restored within the low-energy range). 

For simplicity, the dependencies of the functions and the coefficients
on ${\bf k}$ and the band indices is omitted in these expressions. All
the coefficients are positive. Variation of the matrix elements in
Eqs.~(5--7) is ignored, and sums of the quasiparticle spectral functions
over the band indices and small ${\bf k}^{\prime}$ ranges are expressed
as: 
\begin{eqnarray}
\tilde A^q(\omega) &\cong& \cases{a^q_+\omega + b^q_+ \;,& for
$\omega>0$,\cr -a^q_-\omega + b^q_- \;,& for $\omega<0$,\cr} \\ 
\tilde A^p(\omega) &\cong& \delta(\omega), \\ 
\tilde A^{\zeta}(\omega) &\cong& \cases{a^{\zeta}_+\omega + b^{\zeta}_+
\;,& for $\omega>0$,\cr a^{\zeta}_-\omega - b^{\zeta}_- \;,& for
$\omega<0$.\cr} 
\end{eqnarray} 
These expressions will be derived below self-consistently. By 
Inserting them in Eqs.~(5--7), expressions of the following forms are
derived, assuming the low $T$ limits for $f_{_T}(\omega)$ and
$b_{_T}(\omega)$:
\begin{eqnarray}
{\Gamma^q(\omega) \over 2\pi} &\cong& \cases{c^q_+\omega + d^q_+ \;,&
for $\omega>0$,\cr -c^q_-\omega + d^q_- \;,& for $\omega<0$,\cr} \\ 
{\Gamma^p(\omega) \over 2\pi} &\cong& \cases{c^p_+\omega^3 +
d^p_+\omega^2 + e^p_+\omega \;,& for $\omega>0$,\cr -c^p_-\omega^3 +
d^p_-\omega^2 - e^p_-\omega \;,& for $\omega<0$,\cr} \\ 
{\Gamma^{\zeta}(\omega) \over 2\pi} &\cong& \cases{c^{\zeta}_+\omega +
d^{\zeta}_+ \;,& for $\omega>0$,\cr c^{\zeta}_-\omega - d^{\zeta}_- \;,&
for $\omega<0$.\cr} 
\end{eqnarray} 
Integrating through Eq.~(8) (between the limits $\pm\omega_c$) results
in: 
\begin{eqnarray}
-\Re\Sigma^q&(\omega)& \cong \omega_c(c^q_+ - c^q_-) + \big(d^q_+
\ln{\big|{\omega - \omega_c \over \omega}\big|} \nonumber \\ &-& d^q_-
\ln{\big|{\omega + \omega_c \over \omega}\big|}\big) + \omega\big(c^q_+
\ln{\big|{\omega - \omega_c \over \omega}\big|} \nonumber \\ &+& c^q_-
\ln{\big|{\omega + \omega_c \over \omega}\big|}\big), \\ 
-\Re\Sigma^p&(\omega)& \cong \big[{\omega_c^3 \over 3}(c^p_+ - c^p_-) +
{\omega_c^2 \over 2}(d^p_+ - d^p_-) \nonumber \\ &+& \omega_c(e^p_+ -
e^p_-)\big] + \omega \big[{\omega_c^2 \over 2}(c^p_+ + c^p_-)  \nonumber
\\ &+& \omega_c(d^p_+ + d^p_-) + e^p_+ \ln{\big|{\omega - \omega_c \over
\omega}\big|} \nonumber \\ &+& e^p_- \ln{\big|{\omega + \omega_c \over
\omega}\big|}\big] + \omega^2 \big[\omega_c(c^p_+ - c^p_-) \nonumber \\
&+& d^p_+ \ln{\big|{\omega - \omega_c \over \omega}\big|} - d^p_-
\ln{\big|{\omega + \omega_c \over \omega}\big|}\big] \nonumber \\ &+&
\omega^3 \big[c^p_+ \ln{\big|{\omega - \omega_c \over \omega}\big|} +
c^p_- \ln{\big|{\omega + \omega_c \over \omega}\big|}\big], \\ 
-\Re\Sigma^{\zeta}&(\omega)& \cong \omega_c(c^{\zeta}_+ + c^{\zeta}_-) +
\big(d^{\zeta}_+ \ln{\big|{\omega - \omega_c \over \omega}\big|} 
\nonumber \\ &+& d^{\zeta}_- \ln{\big|{\omega + \omega_c \over
\omega}\big|}\big) + \omega\big(c^{\zeta}_+ \ln{\big|{\omega - \omega_c
\over \omega}\big|} \nonumber \\ &-& c^{\zeta}_- \ln{\big|{\omega +
\omega_c \over \omega}\big|}\big). 
\end{eqnarray} 
The renormalized quasiparticle energies $\bar\epsilon = \epsilon +
\Re\Sigma(\bar\epsilon)$ are obtained by finding the $\omega$ values for
which $\Re\Sigma(\omega) = \omega - \epsilon$. Thus the renormalization
effect of energies well below $\omega_c$ is determined by the slope of
the $\Re\Sigma(\omega)$ {\it vs} $\omega$ curves close to $\omega = 0$. 

The renormalization effect is particularly strong on the stripon
energies. It is contributed by a combined effect of the quasi-continuum
of QE bands within the auxiliary space. This strong effect is reflected
in a significant $\omega^3$ term in $\Gamma^p(\omega)$. Consequently,
the stripon bandwidth drops down to the low energy range, and a
$\delta$-function is appropriate for $\tilde A^p(\omega)$ (11). The
expressions for $\tilde A^q(\omega)$ (10) and $\tilde A^{\zeta}(\omega)$
(12) are obtained by considering the effects of bands crossing zero
energy, which, approximately, contribute constant terms due to the
normalization of the spectral functions (9), and of higher energy bands,
whose contributions are approximately $\propto \Gamma(\omega)$ (9), or
$\propto \omega$. 

The inequality between the coefficients for positive and negative
$\omega$ in Eqs.~(10--15) results mainly because the svivon spectrum is
primarily on the $\omega>0$ side [$\int A^{\zeta}_{\lambda}({\bf k},
\omega) d\omega = 1$] , and the inequality $\cosh{^2(\xi_{{\bf k}})} >
\sinh{^2(\xi_{{\bf k}})}$ of the factors appearing in Eqs.~(5--7). For
the discussed case of p-type cuprates the following inequalities are
built up self-consistently: 
\begin{eqnarray}
a^q_+&>&a^q_-,\ \ \  b^q_+>b^q_-,\ \ \  c^q_+>c^q_-,\ \ \  d^q_+>d^q_-,
\\ a^{\zeta}_+&>&a^{\zeta}_-,\ \ \  b^{\zeta}_+>b^{\zeta}_-,\ \ \
c^{\zeta}_+>c^{\zeta}_-,\ \ \  d^{\zeta}_+>d^{\zeta}_-. 
\end{eqnarray} 
Note, however, that for ``real'' n-type cuprates (namely, ones where the
stripons are based on excession and not holon states) the roles of
$\cosh{(\xi_{{\bf k}})}$ and $\sinh{(\xi_{{\bf k}})}$ are reversed in
${\cal H}^{\prime}$ (4), and the expressions derived from it, and
consequently the direction of the inequalities is reversed for the QE
coefficients (19), but stays the same for the svivon coefficients (20).
Also there could be deviations from (19--20) at specific ${\bf k}$
points, and the inequalities almost disappear for svivons close to point
${\bf k}_0$. 

The spectral functions in Eqs.~(10--12) are within the auxiliary space,
while in order to calculate the physical electrons spectral functions
$A_e({\bf p}, \omega)$, or related physical properties, their projection
into the ``physical space'' is required. Spectroscopies, like ARPES,
measuring the effect of transfer of electrons into, or out of, the
crystal do not detect the low-energy signature of the stripon spectral
functions, because they are smeared over few tenths of an eV through
convolution with svivon spectral functions. Thus they contribute to
$A_e({\bf p}, \omega)$ part of its ``incoherent'' background. On the
other hand, transport properties, discussed below, measure the electrons
{\it within} the crystal, and can detect the stripon energy scale. 

The contribution of the QE spectral functions to $A_e({\bf p}, \omega)$
depends on the expansion coefficients (eigenvectors) of these states in
terms of their basis states [created by (2,3)
$f_{\lambda\sigma}^{\dagger}({\bf k}^{\prime}, {\bf k})$,
$g_{\lambda\sigma}^{\dagger}({\bf k}^{\prime}, {\bf k})$, and
$c_{\nu\sigma}^{\dagger}({\bf k})$]. These eigenvectors have quite 
random phases for almost all the QE states within their quasi-continuum,
and thus contribute to $A_e({\bf p}, \omega)$ another part of its
incoherent background. However, few QE states for which the eigenvectors
closely correspond to those of real electron bands, contribute
``coherent'' $\bar\epsilon^q ({\bf k})$ bands to $A_e({\bf p}, \omega)$.
The occupation factors $n^e_{\lambda}$, $n^h_{\lambda}$, and
$n^s_{\lambda\sigma}$, appearing in Eqs.~(2,3) are reflected in the
observed dependence of $A_e({\bf p}, \omega)$ in correlated electrons
systems on such occupation factors [7]. 

The $\omega$ dependencies (10--20) derived in the auxiliary space remain
valid in the physical space, both for the coherent bands, and for the
incoherent background, detected, {\it e.g.}, in ARPES results. The
present results for $\Gamma^q$ (13) are reflected in the observed
non-Fermi-liquid bandwidths, having a $\propto\omega$ and a constant
term. The results for $\Re\Sigma^q$ (16) are reflected in band slopes
which are smaller than the LDA predictions. Also the fact
that the QE field is coupled to the stripon and svivon fields within a
BZ corresponding to their (lower) periodicities is experimentally
reflected in ``shadow bands'' and further features associated with a
superstructure. 

There is a peculiar consequence to the logarithmic singularity in
$\Re\Sigma^q$ at $\omega=0$, due to the $(d^q_+ - d^q_-)\ln{|\omega|}$
term in Eq.~(16). For p-type cuprates (19) the consequence of this
singularity is that as an $\bar\epsilon^q({\bf k})$ band is getting close
to $E_{_{\rm F}}$ from below it is becoming flatter (though it does not
narrow accordingly) while from above it becomes steeper, and may pass
through an infinite slope resulting in an S-shape triply valued band
(this causes smearing of the spectral weight, and blurring the band;
note, however, that the effect of the singularity is truncated and
smoothened on an energy scale $\lta 0.02\;$eV). For ``real'' n-type
cuprates these behaviors below and above $E_{_{\rm F}}$ are switched.
ARPES data reflects the behavior below $E_{_{\rm F}}$, and results of
such band-flattening have been reported in p-type cuprates (recently
such an effect has been attributed to electron coupling with phonons [8]
or with the resonance mode observed in neutron scattering [9]; also the
observed renormalization of van Hove singularities to extended ones may
reflect this effect). The behavior above $E_{_{\rm F}}$ would be
detected in ARIPES measurements. Careful measurements of the bands of
n-type cuprates very close to $E_{_{\rm F}}$ would help determine
whether they are ``real'' n-type ones. 

$\Re\Sigma^{\zeta}(\omega)$ (18) is negative in the intermediary energy
range, and its $\omega$ dependence is weak, except for the vicinity of
$\omega=0$, where a negative logarithmic singularity exists, due to 
the $(d^{\zeta}_+ + d^{\zeta}_-)\ln{|\omega|}$ term in Eq.~(18), turning
into a smooth minimum in the low energy range. Consequently, the svivon
quasiparticle energies are renormalized through a negative energy shift,
resulting in negative $\bar\epsilon^{\zeta}({\bf k})$ values close to
their minimum at ${\bf k}_0$. This does not introduce divergences at the
low energy range, since in this scale $A^{\zeta}(\omega)$ is analytic
and $\propto \omega$ around $\omega=0$. 

The effect of the logarithmic singularity, and its smoothening at low
energy, is that as ${\bf k}_0$ is approached, the slope of
$\bar\epsilon^{\zeta}({\bf k})$ increases first [compared to the almost
constant slope of $\epsilon^{\zeta}({\bf k)}$] and it may become
infinite, resulting in a range of an S-shape triply valued and blurred
band. But as ${\bf k}_0$ is further approached, the slope of
$\bar\epsilon^{\zeta}({\bf k})$ decreases and becomes smaller than the
slope of $\epsilon^{\zeta}({\bf k})$ at its minimum at ${\bf k}_0$. It
is likely (though we cannot prove it under the present approximations)
that the slope in this minimum is zero, and that
$\bar\epsilon^{\zeta}({\bf k})$ has an analytic minimum [rather than the
V-shape minimum of $\epsilon^{\zeta}({\bf k})$], consistently with the
fact that there is no long-range AF order. Thus
$-\bar\epsilon^{\zeta}({\bf k}_0)$ presents the crossover energy from
the intermediary to the low energy range. Spin-flip excitations are,
largely, double-svivon absorption or emission processes. Thus,
$-2\bar\epsilon^{\zeta}({\bf k}_0)$ is the energy of such an excitation
at ${\bf k} = {\bf Q}$. This is consistent with the neutron-scattering
resonance energy observed at this wave number [10]. The observation that
this resonance energy is at a local maximum (for ${\bf k}$ around ${\bf
Q}$) is consistent with the present result that
$\bar\epsilon^{\zeta}({\bf k}_0)$ is at a minimum below zero. 

The optical conductivity of the doped cuprates is characterized [11] by
an anomalous Drude term and a mid-IR term. The Drude term results from
transitions between QE states, and (at low $\omega$) also between
stripon states. The mid-IR term results (at least partly) from
transitions between stripon states and QE+svivon states. The mid-IR term
becomes negligibly small for $\omega \to 0$, and one can decouple
between the contributions of the QE's and stripons to conductivity for
$\omega \to 0$. 



Thus, low-$\omega$ electric current can be expressed as a sum: ${\bf j}
= {\bf j}^q + {\bf j}^p$, of QE and stripon contributions (the svivons
do not carry electric charge, and their effect on the current is through
occupation factors).  Since both the QE and the stripon states are
expressed through a perturbative expansion in the auxiliary space in
terms of ``bare'' QE and and stripon states, one can also express ${\bf
j} = {\bf j}^q_0 + {\bf j}^p_0$, where ${\bf j}^q_0$ and ${\bf j}^p_0$
are the contributions of these bare states. However, since the bare
stripon states are localized, one can assume ${\bf j}^p_0 \cong 0$, and
thus ${\bf j} \cong {\bf j}^q_0$ (note that ${\bf j}^q_0$ and ${\bf
j}^p_0$ depend on the velocity distribution of the bare states, and
localized states correspond to an infinite ``mass'' and zero velocity).
Consequently one can express: 
\begin{equation}
{\bf j}^q / (1-\alpha) \cong {\bf j}^p / \alpha \cong {\bf j}^q_0 \cong
{\bf j}, 
\end{equation}
where $\alpha$ depends on expansion coefficients and occupation factors,
and thus has a negligible temperature dependence. 

When an electric field is applied, condition (21) is satisfied by the
formation of gradients ${\bfnabla}\mu^q$ and ${\bfnabla}\mu^p$ of the QE
and stripon chemical potentials, respectively (the treatment of the
constraint provides an additional Lagrange multiplier to the electrons
chemical potential). Charge neutrality imposes: 
\begin{equation}
{\partial n^q_e \over \partial \mu^q} {\bfnabla}\mu^q + {\partial n^p_e
\over \partial \mu^p} {\bfnabla}\mu^p =0. 
\end{equation}
where $n^q_e$ and $n^p_e$ are the contributions of QE and stripon states
to the electrons occupation. If one introduces: 
\begin{equation}
N^q_e \equiv {\partial n^q_e \over \partial \mu^q}, \ \ \ \ \ M^p_e(T)
\equiv T {\partial n^p_e \over \partial \mu^p}, 
\end{equation}
it turns out, due to the QE and stripon bandwidths, that $N^q_e$ is
almost temperature independent, and that $M^p_e(T)$ is temperature
dependent only at low $T$. In the presence of an electric field ${\bf
E}$ one can express: 
\begin{equation}
{\bfcalE}^q =  \rho^q(T) {\bf j}^q, \ \ \ \ \ {\bfcalE}^p =  \rho^p(T)
{\bf j}^p 
\end{equation}
where
\begin{equation}
{\bfcalE}^q={\bf E} + {\bfnabla}\mu^q / {\rm e}, \ \ \ \ \
{\bfcalE}^p={\bf E} + {\bfnabla}\mu^p / {\rm e}, 
\end{equation}
and by Eqs.~(22,23,25): 
\begin{equation}
{\bf E}={M^p_e(T) {\bfcalE}^p + N^q_e {\bfcalE}^q T \over M^p_e(T) +
N^q_eT}.
\end{equation}
Since ${\bf E} = \rho {\bf j}$, the electrical resistivity $\rho$ can be
expressed, using Eqs.~(21,24,26), as: 
\begin{equation}
\rho ={\alpha M^p_e(T) \rho^p(T) + (1 - \alpha) N^q_e \rho^q(T) T \over
M^p_e(T) + N^q_eT}. 
\end{equation}
This expression demonstrates that the anomalous temperature dependence
of $\rho$ is determined here by $\rho^q(T)$, $\rho^p(T)$, and also by
the temperature dependence of $\partial n^p_e / \partial \mu^p =
M^p_e(T)/T$. Expressions will be derived below on the basis of a band
model. 

Eq.~(21) also plays an important role in the determination of the Hall
constant $R_{_{\rm H}} = 1/{\rm e} n_{_{\rm H}}$. When a magnetic field
$H = H_z$ is applied under current $j_x$, and an electric field $E_y$ is
required to keep $j_y =0$, then the ``Hall number'' is: $n_{_{\rm H}} =
j_xH / {\rm e} E_y$. However, by Eq.~(21) one then gets also $j^q_y =
j^p_y =0$, and QE and stripon ``Hall effects'' can be treated separately ,
yielding: 
\begin{equation}
{\cal E}^q_y= j^q_xH / {\rm e} n^q_{_{\rm H}}, \ \ \ \ \ {\cal E}^p_y=
j^p_xH / {\rm e} n^p_{_{\rm H}}, 
\end{equation}
where $n^q_{_{\rm H}}$ and $n^p_{_{\rm H}}$ can be roughly interpreted
as effective QE and stripon contributions to the density of charge
carriers, and are not expected to be temperature dependent. Since (21)
both ${\bf j}^q$ and ${\bf j}^p$ are determined by the current ${\bf
j}^q_0$ of the bare QE states, both $n^q_{_{\rm H}}$ and $n^p_{_{\rm
H}}$ are expected to have same sign which corresponds to these states.
By Eqs.~(26,28) one can then express the Hall number as: 
\begin{equation}
n_{_{\rm H}} = {M^p_e(T) + N^q_eT \over \alpha M^p_e(T) / n^p_{_{\rm H}}
+ (1 - \alpha) N^q_e T / n^q_{_{\rm H}}}. 
\end{equation}
Here, the anomalous temperature dependence is determined only by
$M^p_e(T)/T$. A quantity often discussed is the Hall angle
$\theta_{_{\rm H}}$, defined through: $\cot{\theta_{_{\rm H}}} = \rho /
R_{_{\rm H}}$. By Eqs.~(27,29) it is expressed as: 
\begin{equation}
\cot{\theta_{_{\rm H}}} = {{\rm e} [\alpha M^p_e(T) \rho^p(T) + (1 -
\alpha) N^q_e \rho^q(T) T] \over  \alpha M^p_e(T) / n^p_{_{\rm H}} + (1
- \alpha) N^q_e T / n^q_{_{\rm H}}}. 
\end{equation}

When both an electric field and a temperature gradient are present, 
one can express ${\bf j}^q$ and ${\bf j}^p$ as:
\begin{eqnarray}
{\bf j}^q &=& {{\rm e} \over T} L^{q(11)} {\bfcalE}^q + L^{q(12)}
{\bfnabla}({1 \over T}), \\ {\bf j}^p &=& {{\rm e} \over T} L^{p(11)}
{\bfcalE}^p + L^{p(12)} {\bfnabla}({1 \over T}). 
\end{eqnarray} 
These expressions are used to evaluate the 
thermoelectric power (TEP), defined as: $S = [E / {\nabla}
T]_{{\bf j} = 0}$. Since (21) the condition ${\bf j}=0$ corresponds also
to ${\bf j}^q = {\bf j}^p = 0$, one can get by Eqs.~(31,32) expressions
for QE and stripon ``TEP's'': 
\begin{eqnarray}
S^q(T) &=& \bigg[{{\cal E}^q \over {\nabla} T}\bigg]_{{\bf j}^q = 0} = -
{1 \over {\rm e}T} { L^{q(12)} \over L^{q(11)}}, \\ S^p(T) &=&
\bigg[{{\cal E}^p \over {\nabla} T}\bigg]_{{\bf j}^p = 0} = - {1 \over
{\rm e}T} {L^{p(12)} \over L^{p(11)}}, 
\end{eqnarray}
and through Eq.~(26) express the TEP as: 
\begin{equation}
S={M^p_e(T) S^p(T) + N^q_e S^q(T)T \over M^p_e(T) + N^q_eT}. 
\end{equation}

In order to get the temperature dependencies in the above transport
expressions we need expressions for $\tilde A^q(\omega)$, $\tilde
A^p(\omega)$, and $\tilde A^{\zeta}(\omega)$ in the low energy range
[rather than the intermediary energy range expressions (10--12)]. As was
mentioned above, the non-analytic behavior of the spectral functions
disappears within the low energy range. For $\tilde A^q$ and $\tilde
A^{\zeta}$ we assume the lowest order terms in a series expansion in
$\omega$, thus: 
\begin{equation}
\tilde A^q(\omega) \cong {\rm const}, \ \ \ \ \ \tilde A^{\zeta}(\omega)
\propto \omega. 
\end{equation}
A band model of a ``rectangular" shape of width $\omega^p$, and
fractional occupation $n^p$, is assumed for $\tilde A^p$: 
\begin{equation}
\tilde A^p(\omega) = \cases{{1 \over \omega^p} \;,& for $ -{\omega^p
\over 2} < \omega + \mu < {\omega^p \over 2}$,\cr 0 \;,& otherwise,\cr} 
\end{equation}
where $\mu \equiv \mu^p$ (its position serves as the energy zero). Note
that the inclusion of temperature dependencies in Eqs.~(36,37), as well
as terms of higher $\omega$ powers in Eq.~(36), would add terms of
higher powers of $T$ to the results, and thus have an effect at
high-temperatures. 

Since $n^p = \int \tilde A^p(\omega) f_{_T}(\omega) d\omega$, it is
straight forward to express $\mu$ and $M^p_e$ [we use  Eq.~(23),
normalize the occupation numbers such that $n^p_e = n^p$, and express
$T$ in energy units] as: 
\begin{eqnarray}
\mu &=& T\ln{\big\{}\{\exp{[n^p\omega^p / 2T]} - \exp{[-n^p\omega^p /
2T]} \} \nonumber \\ &/& \{\exp{[(1-n^p)\omega^p / 2T]} \nonumber \\ &-&
\exp{[-(1-n^p)\omega^p / 2T]} \} \big\}, \\ 
M^p_e &=& (T / \omega^p)[f_{_T}(-\half\omega^p - \mu) -
f_{_T}(\half\omega^p - \mu)]. 
\end{eqnarray} 
Their low and high $T$ limits are given by:
\begin{eqnarray}
\mu &=& \cases{(n^p - \half) \omega^p \;,& for $T \ll \omega^p$,\cr
T\ln{[n^p/(1-n^p)]} \;,& for $T \gg \omega^p$,\cr} \\ 
M^p_e &=& \cases{T/\omega^p \;,& for $T \ll \omega^p$,\cr n^p(1-n^p)
\;,& for $T \gg \omega^p$.\cr} 
\end{eqnarray} 

Expressions for $\rho^p(T)$ and $\rho^q(T)$ (24) are derived using
linear response theory, under which $\rho^p(T) \propto \Gamma^p(T)$ and
$\rho^q(T) \propto \Gamma^q(T)$. Eqs.~(5,6) are applied to evaluate
$\Gamma^p(\omega=0)$ and $\Gamma^q(\omega=0)$, and temperature
independent terms are added to account for impurity scattering. Thus:
\begin{eqnarray}
\Gamma^p &\propto& \int \tilde A^q(\omega) \tilde A^{\zeta}(\omega)
[f_{_T}(\omega) + b_{_T}(\omega)] d\omega + {\rm const}, \\
\Gamma^q &\propto& \int \tilde A^p(\omega) \tilde A^{\zeta}(\omega)
[f_{_T}(\omega) + b_{_T}(\omega)] d\omega + {\rm const},
\end{eqnarray}
where the model spectral functions (36,37) are used. The result for
$\Gamma^p$ is straight forward, yielding: 
\begin{equation}
\rho^p \cong \rho^p_0 + \rho^p_2 T^2. 
\end{equation}
The integral (43) for $\Gamma^q$ does not have a rigorous analytical
result, however an approximate one can be derived based on the two
leading terms in a high-$T$ expansion, joined smoothly with an
approximate low-$T$ limit term. It can be expressed as: 
\begin{equation}
\rho^q \cong \rho^q_0 + \rho^q_1\cases{AT - {B \over T} \;,& for $T >
\sqrt{2B \over A}$,\cr {A \over 2}\sqrt{A \over 2B}T^2 \;,& for $T <
\sqrt{2B \over A}$,\cr} 
\end{equation}
where the positive constants $A$ and $B$ are expressed in terms of the
above stripon band parameters. 

The stripons contribution to the TEP is derived here through [12]: 
\begin{equation}
S^p \cong -{k_{_{\rm B}} \over {\rm e}} {\int  \tilde A^p(\omega)
{\omega \over T} {df_{_T}(\omega) \over d\omega} d\omega \over \int
\tilde A^p(\omega) {df_{_T}(\omega) \over d\omega} d\omega}, 
\end{equation}
where $k_{_{\rm B}}$ is Boltzmann's constant ($k_{_{\rm B}} / {\rm
e} = 86 \; \mu$V/K). Inserting (37) in (46), and using (38,39), one 
gets:
\begin{eqnarray}
S^p \cong {k_{_{\rm B}} \over {\rm e} M^p_e} &\Big\{\half&
[f_{_T}(-\half\omega^p - \mu) + f_{_T}(\half\omega^p - \mu)] \nonumber
\\ &+& {\mu M^p_e \over T} -n^p \Big\}, 
\end{eqnarray}
with low and high $T$ limits:
\begin{equation}
S^p \cong {k_{_{\rm B}} \over {\rm e}} \cases{ 0\;,& for $T \ll
\omega^p$,\cr \ln{[n^p/(1-n^p)]} \;,& for $T \gg \omega^p$.\cr}
\end{equation}

In order to apply an expression similar to Eq.~(46) to derive $S^q$
(replacing $\tilde A^p$ there by $\tilde A^q$), a linear term $a^q
\omega$ should be added to $\tilde A^q(\omega)$ in Eq.~(36), resulting
in a metallic linear $T$ behavior: 
\begin{equation}
S^q \cong S^q_1 T, 
\end{equation}
where the constant $S^q_1$ has an opposite sign than $a^q$. The
inequalities (19), and the smoothening of the discontinuity in $\tilde
A^q(\omega)$ in the low energy range, results in $a^q>0$, and thus
$S^q_1<0$ for p-type cuprates [for which (19) is valid] while [following
the discussion after Eq.~(19)] for ``real'' n-type curates $a^q<0$, and
thus $S^q_1>0$. 

The transport coefficients are calculated using
Eqs.~(27,29,30,35,38,39,44,45,47,49) for five stoichiometries, ranging
from underdoped ($n^p=0.8$) to overdoped ($n^p=0.4$) p-type cuprates.
$N^q_e$ is assumed to increase with doping, reflecting transfer of QE
spectral weight towards $E_{_{\rm F}}$. Consequently (6) $\omega^p$ is
assumed to decrease with doping. We introduce $\tilde n^p_{_{\rm H}}
\equiv n^p_{_{\rm H}}/\alpha$ and $\tilde n^q_{_{\rm H}} \equiv
n^q_{_{\rm H}}/(1 - \alpha)$, and they are assumed to increase with
doping, where $\tilde n^p_{_{\rm H}}$ increases considerably faster than
$\tilde n^p_{_{\rm H}}$, and a little faster than $1-n^p$ (since $\tilde
n^p_{_{\rm H}}$ reflects an overall stripon-related carriers density,
while $n^p$ is the fractional occupation of stripon states within the
charged stripes whose number increases with doping). In order to
separate the effect of doping on the density of states and the carriers
density we introduce $\tilde \rho^p_0 \equiv \rho^p_0 n^p_{_{\rm H}} /
N^q_e$, $\tilde \rho^p_2 \equiv \rho^p_2 n^p_{_{\rm H}} / N^q_e$,
$\tilde \rho^q_0 \equiv \rho^q_0 n^q_{_{\rm H}} (1-n^p) / n^p_{_{\rm
H}}$, and $\tilde \rho^q_1 \equiv \rho^q_1 n^q_{_{\rm H}} (1-n^p) /
n^p_{_{\rm H}}$, and assume doping-independent values of $\tilde
\rho^p_0$, $\tilde \rho^p_2$, $\tilde \rho^q_0$, and $\tilde \rho^q_1$.
Results for a specific choice of the parameters are presented in Fig.~1.
These results, however, do {\it not} include the effect of the
pseudogap. 

The results for $S$ [Fig.~1(a)] reproduce very well the systematic TEP
behavior, which is characteristic of the doping level [13--15]
(determined here mainly through $n^p$). The position of the maximum in
$S$ depends on the choice of $\omega^p$. This maximum may be below or
above $T_c$, and the occurrence of a pseudogap may shift it to a higher
temperature than predicted here. Electron bands which are not coupled to
the stripons in the CuO$_2$ planes may have a linear contribution to $S$
deviating from our prediction. Also the results for $n_{_{\rm H}}$ and
$R_{_{\rm H}}$ [Fig.~1(b,c)] reproduce very well the systematic Hall
behavior [16,17]. The increasing linear $T$ dependence of $n_{_{\rm H}}$
reflects a crossover from, approximately, the smaller $\tilde n^p_{_{\rm
H}}$ at low $T$ to the larger $\tilde n^q_{_{\rm H}}$ at high $T$. The
results for $\rho$ and $\cot{\theta_{_{\rm H}}}$ [Fig.~1(d,e)] reproduce
very well the systematic experimental behavior too [16,17]. An
approximate quadratic $T$ dependence of $\cot{\theta_{_{\rm H}}}$ is
obtained in the temperature range where $n_{_{\rm H}}$ is linear in $T$.
The linear $T$ behavior of $\rho$ persists through the high-temperature
saturation range of $R_{_{\rm H}}$ [17]. Cases of non-quadratic $T$
dependence of $\rho$ at very low $T$ are expected as a pseudogap effect.

In ``real'' n-type cuprates $S$ is expected to behave similarly to p-type
cuprates, for similar doping levels, but with an opposite sign (and
slope). TEP results for NCCO [18] show that in superconducting doping
levels it has the same behavior as in p-type \eject 

\begin{figure}[t]
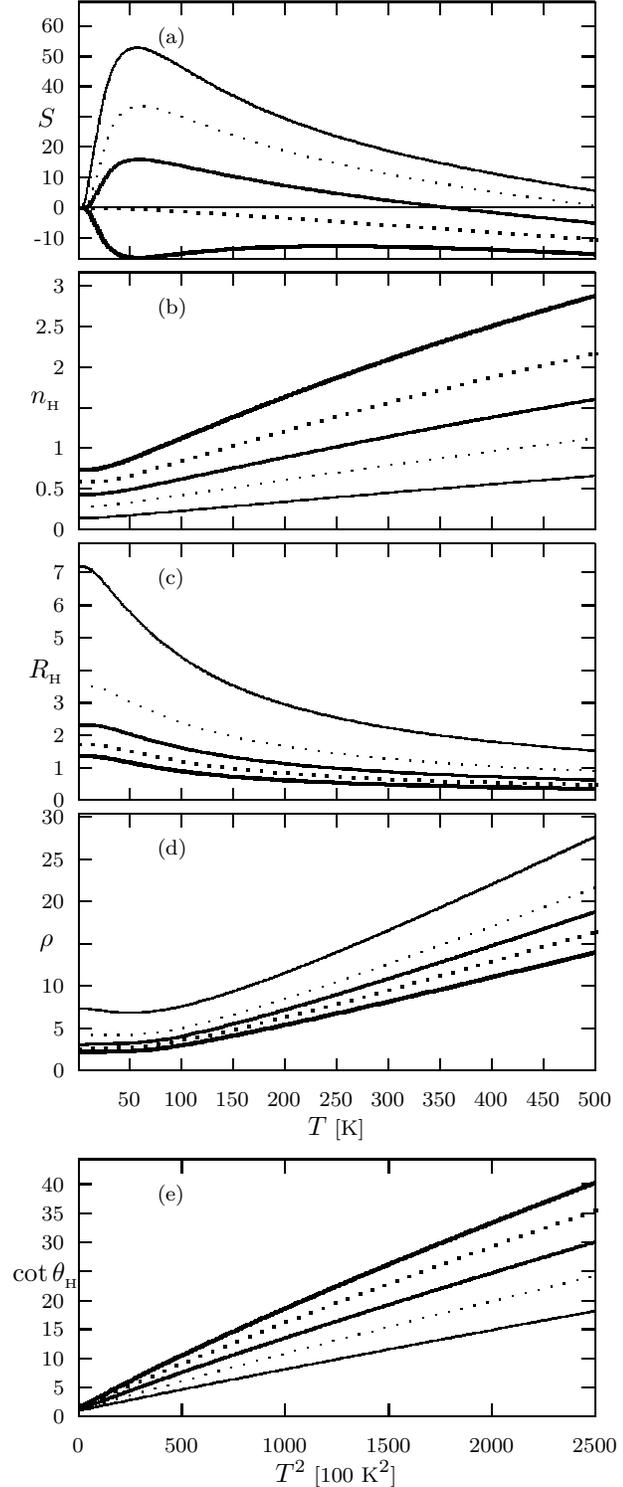


\setlength{\unitlength}{0.240900pt}
\ifx\plotpoint\undefined\newsavebox{\plotpoint}\fi
\sbox{\plotpoint}{\rule[-0.200pt]{0.400pt}{0.400pt}}%


\caption{The transport coefficients, in arbitrary units [and $\mu$V/K
units for $S$ (a)], for: $n^p$=0.8,0.7,0.6,0.5,0.4;
$10000N^q_e$=20,23,26,29,32; $\omega^p$[K]=200,190,180,170,160;
$\tilde n^p_{_{\rm H}}$=0.1,0.2,0.3,0.4,0.5; $\tilde n^q_{_{\rm
H}}$=6,7,8,9,10; $S^q_1$=$-$0.025; $\tilde \rho^p_0$=500; $\tilde
\rho^p_2$=0.03; $\tilde \rho^q_0$=5; $\tilde \rho^q_1$=0.2. The last
values correspond to the thickest lines.} 
\label{F1}
\end{figure}

\noindent cuprates [Fig.~1(a)], but with an opposite effect of doping.
Namely $S$ for lower doping levels is typical of the p-type overdoped
regime. This indicates that NCCO is probably not a ``real'' n-type
cuprate, and its stripons are also based on holon states, where the
extra doped negative charge occupies another QE band. The Hall constant
in NCCO [18] is consistent with Fig.~1(d), and our prediction that the
sign of both $n^p_{_{\rm H}}$ and $n^q_{_{\rm H}}$ is determined by the
bare QE states (and changes in NCCO with doping). 

The coupling Hamiltonian ${\cal H}^{\prime}$ (4) provides a mechanism
for high-$T_c$. The pairing is introduced by transitions between pair
states of stripons and QE's through the exchange of svivons [5]. Such
transitions enable long-range hopping of local stripon pairs, without an
associated hopping of svivons (necessary for single-stripon hopping).
Thus the pairing results in a large gain in the ``kinetic'' energy of
the stripons, which can effectively be expressed as a strong attractive
stripon-stripon interaction. 

Pair-breaking is expected here to result in QE and stripon+svivon
excitations. ARPES measurements in the superconducting state [19] show a
sharp peak at $\sim 0.04\;$eV over a wide range of the BZ, and a
``hump'' starting at $\sim 0.1\;$eV and merging with a normal-state band
at higher energies. The hump is consistent with a QE pair-breaking
excitation, and the peak is consistent with a stripon+svivon
pair-breaking excitation at the svivon minimum excitation energy
$-\bar\epsilon^{\zeta}({\bf k}_0)$. In the superconducting state this
minimum is within the gap, resulting in a considerably narrower level
than in the normal state [reflected also in the neutron-scattering
resonance energy $-2\bar\epsilon^{\zeta}({\bf k}_0)$]. This results in a
sharp pair-breaking excitation peak over a wide range of the BZ, due to
the flatness of the stripon band. 

It has been suggested [20] that the cuprates fall in the regime of 
crossover between BCS and  preformed-pairs Bose-Einstein condensation
(BEC) behaviors. Such a crossover occurs if the pairing energy becomes
as large as the relevant bandwidth, and this condition is fulfilled here
due to the small stripon bandwidth. The BEC behavior occurs in the
underdoped cuprates, where singlet pairs are formed at $T_{\rm pair}$,
and superconductivity occurs below $T_{\rm coh} (< T_{\rm pair})$ where
phase coherence sets in. 

The pseudogap is then a pair-breaking gap between $T_{\rm coh}$ and
$T_{\rm pair}$; it has similarities to the superconducting gap, and
accounts for most of the pairing energy. The coherence temperature can
be expressed [21] as: $T_{\rm coh} \propto n_s / m_s^*$, where $m_s^*$
in the pairs effective mass and $n_s$ is their density, in agreement
with the ``Uemura plots'' [22]. Our observation [Fig.~1(a)] that the
stripon band if half full ($n^p=\half$) for slightly overdoped cuprates
results in the ``boomerang-type'' behavior [23] of the Uemura plots in
overdoped cuprates. It is due to a crossover between a stripon band-top
$T_c = T_{\rm coh}$ behavior in the underdoped cuprates, and a stripon
band-bottom $T_c = T_{\rm pair}$ behavior in overdoped cuprates. 


In conclusion, it was demonstrated here that a realistic treatment of
the cuprates, taking into account the effects of both large-$U$ and
small-$U$ orbitals, and of the inhomogeneities introduced by a dynamical
striped structure, resolves puzzling normal-state spectroscopic and
transport properties, and provides a mechanism for high-$T_c$
superconductivity and the occurrence of a normal-state pseudogap. 

\bigskip
\centerline{\bf REFERENCES} 
\medskip
\baselineskip 10pt
\footnotesize
\noindent
1. \ S.~E.~Barnes, {\it Adv.~Phys.} {\bf 30}, 801 (1980). \\
2. \ J.~Ashkenazi, {\it J.~Supercond.} {\bf 7}, 719 (1994). \\
3. \ V.~J.~Emery, and S.~A.~Kivelson, {\it Physica C} {\bf 209}, 597 
\\ \mbox{\ } \ \ \ (1993). \\
4. \ Papers in {\it Int.~J.~Mod.~Phys.} {\bf 14}, 3289--3790 (2000). \\
5. \ J.~Ashkenazi, {\it High-Temperature Superconductivity}, \\ \mbox{\
} \ \ \ edited by S.~E.~Barnes, J.~Ashkenazi, J.~L.~Cohn, and \\ \mbox{\
} \ \ \ F.~Zuo (AIP Conference Proceedings 483, 1999), p. 12; \\ \mbox{\
} \ \ \ cond-mat/9905172. \\
6. \ A.~Bianconi, {\it et al.}, {\it Phys.~Rev.~B} {\bf 54}, 12018 (1996).
\\ 
7. \ H.~Eskes, {\it et al.}, {\it Phys. Rev. Lett.} {\bf 67}, 1035
(1991). \\ 
8. \ Z.~X.~Shen, {\it et al.}, cond-mat/0108381; this issue. \\
9. \ P.~D.~Johnson, {\it et al.}, {\it Phys. Rev. Lett.} {\bf 87},
177007 \\ \mbox{\ } \ \ \ (2001). \\ 
10.  H.~F.~Fong, {\it et al.}, {\it Phys.~Rev.~B} {\bf 61}, 14773
(2000). \\ 
11.  D.~B.~Tanner, and T.~Timusk, {\it Physical Properties \\ \mbox{\ }
\ \ \ of High Temperature Superconductors III}, edited by \\ \mbox{\ } \
\ \ D.~M.~Ginsberg (World Scientific, 1992), p. 363. \\ 
12.  S.~Bar-Ad, B. Fisher, J. Ashkenazi and J. Genossar, \\ \mbox{\ } \
\ \ {\it Physica C} {\bf 156}, 741 (1988). \\ 
13. B.~Fisher, {\it et al.}, {\it J. Supercond.} {\bf 1}, 53 (1988);
J. Genossar, \\ \mbox{\ } \ \ \ {\it et al.}, {\it Physica C} {\bf
157}, 320 (1989). \\ 
14. S. Tanaka, {\it et al.}, {\it J.~Phys.~Soc.~Japan} {\bf 61}, 1271 
(1992). \\
15. K.~Matsuura, {\it et al.}, {\it Phys.~Rev.~B} {\bf 46}, 11923
(1992); \\ \mbox{\ } \ \ \ S.~D.~Obertelli, {\it et al.}, {\it ibid.},
p. 14928; \\ \mbox{\ } \ \ \ C.~K.~Subramaniam, {\it et al.}, {\it
Physica C} {\bf 203}, 298 (1992). \\ 
16. Y.~Kubo and T.~Manako, {\it Physica C} {\bf 197}, 378 (1992). \\ 
17. H.~Takagi, {\it et al.}, {\it Phys.~Rev.~Lett.} {\bf 69}, 2975
(1992); \\ \mbox{\ } \ \ \ H.~Y.~Hwang, {\it et al.}, {\it ibid.} {\bf
72}, 2636 (1994). \\ 
18. J.~Takeda, {\it et al.}, {\it Physica C} {\bf 231}, 293
(1994); X.-Q.~Xu, \\ \mbox{\ } \ \ \ {\it et al.}, {\it Phys.~Rev.~B}
{\bf 45}, 7356 (1992); Wu Jiang, {\it et al.}, \\ \mbox{\ } \ \ \ {\it
Phys.~Rev.~Lett.} {\bf 73}, 1291 (1994). \\ 
19. M.~R.~Norman, and H.~Ding, {\it Phys.~Rev.~B} {\bf 57}, R11089 \\
\mbox{\ } \ \ \ (1998). \\ 
20. M.~Randeria, cond-mat/9710223, Varenna Lectures \\ \mbox{\ } \ \ \
(1997); J.~R.~Engelbrecht, {\it et al.}, {\it Phys.~Rev.~B} {\bf 55},
15153 \\ \mbox{\ } \ \ \ (1997); R.~D.~Duncan, and
C.~A.~R.~S\'a~de~Melo, {\it ibid} {\bf 62}, \\ \mbox{\ } \ \ \ 9675
(2000); Q.~Chen, {\it et al.}, {\it ibid} {\bf 63}, 184159 (2001). \\ 
21. V.~J.~Emery, and S.~A.~Kivelson, {\it Nature} {\bf 374}, 4347 \\
\mbox{\ } \ \ \ (1995). \\ 
22. Y.~J.~Uemura, {\it et al.}, {\it Phys.~Rev.~Lett.} {\bf 62}, 2317
(1989). \\ 
23. Ch.~Niedermayer, {\it et al.}, {\it Phys.~Rev.~Lett.}~{\bf 71}, 1764
\\ \mbox{\ } \ \ \ (1993). \\ 

\end{document}